\documentstyle[12pt]{article}
\newcommand{\be}{\begin{equation}}
\newcommand{\ee}{\end{equation}}
\newcommand{\bea}{\begin{eqnarray}}
\newcommand{\eea}{\end{eqnarray}}
\newcommand{\bean}{\begin{eqnarray*}}
\newcommand{\eean}{\end{eqnarray*}}
\newcommand{\ba}{\begin{array}}
\newcommand{\ea}{\end{array}}

\newcommand{\norsl}{\normalsize\sl}
\newcommand{\norsc}{\normalsize\sc}
\begin{document}

\begin{titlepage}

\title
{Does Leading $\ln x$ Resummation Predict the Rise
     of $g_1$ at Small $x$ ?}

\author{
\norsc  Yuichiro KIYO\thanks{E-mail :
kiyo@theo.phys.sci.hiroshima-u.ac.jp}, Jiro KODAIRA\thanks{E-mail :
kodaira@theo.phys.sci.hiroshima-u.ac.jp} and Hiroshi TOCHIMURA\thanks{E-mail :
tochi@theo.phys.sci.hiroshima-u.ac.jp}\\
\norsl  Dept. of Physics, Hiroshima University\\
\norsl  Higashi-Hiroshima 739, JAPAN\\
\\
}

\date{}

\maketitle

\begin{abstract}
{\normalsize 
We numerically analyse the evolution of the flavor non-singlet
$g_{1}$ structure function taking into account the
all-order resummation of $\alpha_{s} \ln^{2}x $ terms which is expected
to have much stronger effects than the DGLAP evolution
in the small $x$ region.
We include a part of the next-to-leading logarithmic corrections
coming from the resummed \lq\lq coefficient function\rq\rq\ which
are not considered in the calculation of 
Bl\"umlein and Vogt to respect
the factorization scheme independence.
It is pointed out that the resummed coefficient function
gives unexpectedly large suppression factor over the
experimentally accessible range of $x$ and $Q^{2}$.
This fact implies that the next-to-leading logarithmic
contributions are very important for the $g_{1}$ structure function.
}
\end{abstract}

\begin{picture}(5,2)(-330,-480)
\put(2.3,-65){HUPD-9618}
\end{picture}
 
\vspace{2cm}
\leftline{\hspace{1.2cm}hep-ph/9701365}

\leftline{\hspace{1.2cm}December 1996}

\thispagestyle{empty}
\end{titlepage}
\setcounter{page}{1}
\baselineskip 24pt

\section{Introduction}

Recent new data for the unpolarized 
Deep Inelastic Scattering from HERA
provide us with much information on the quark-gluon structure of Nucleon.
The HERA experiments cover much broader kinematical regions
than before.
Especially, the behavior of the structure function
at small values of the Bjorken variable $x$ receives much attention
of the physicists~\cite{intro}.
The small $x$ region corresponds to the Regge limit.
So we naively expect that the soft physics (Regge theory) may
explain the small $x$ behavior of the structure function.
However the steep rise of the structure function in this
region observed by the HERA experiments contradicts with this naive
expectation. 
The physics at small $x$ is now one of the most interesting
subjects
and many people believe that this problem could be handled in the context
of the QCD perturbation theory~\cite{intro}.
The approach based on the Balitskii-Fadin-Kuraev-Lipatov
(BFKL)~\cite{bfkl} equation or on the high-energy
factorization~\cite{catani}
seems to be very promising and it is important to reveal the role of the
so-called BFKL Pomeron.

In the case of the polarized structure function $g_{1}$,
we have not yet had data at very small $x$.
However the recent data show some rise of $g_1$
in the small $x$ region~\cite{smc}.
This behavior again seems to contradict with naive Regge prediction
($ g_{1} \sim x^{\alpha}  , \  0 \leq \alpha \leq 0.5 $)~\cite{heimann}.
In fact, to explain the rise of $g_{1}$ at small $x$ in the
framework of the Dokshitzer-Gribov-Lipatov-Altarelli-Parisi
(DGLAP)~\cite{dglap} approach, it is required to choose a steep function
as an input parton density~\cite{bfr}
as far as the $Q^{2}$ evolution starts at the order of
1 $GeV^2$ because the evolution
effect in the DGLAP equation does not produce enough
enhancement in the small $x$ region~\cite{grsv}.
So it is interesting to see whether the all-order resummation of 
$\ln x $ terms which appear in the perturbative calculations
reconcile the experimental behavior of $g_{1}$ with
the naive Regge prediction.

Some time ago, Kirschner and Lipatov~\cite{kili} considered the
all order resummation of $\alpha_{s} \ln^{2}x $ series
in the case of quark-quark forward scattering process.
Recently Bartels {\sl et al}.~\cite{bartels} have given
the resummed expression for the
$g_{1}$ structure function by using
the Infra-Red Evolution Equation.
They claim that the resummation effects may lead to 10 times larger
results than  the DGLAP ones.
This analysis suggests the above possibility that the small $x$ behavior
of $g_{1}$ is explained naturally
by combining a flat input (non-perturbative) density  expected
from the naive Regge theory (it is reasonable at low $Q^{2}$,$x$) 
with the perturbatively resummed results of $\alpha_{s} \ln^{2}x $
series.

On the other hand, the recent numerical analysis which has
been done by 
Bl\"umlein and Vogt~\cite{blvo} shows that there are no significant
contributions to the evolution of $g_1$ from the resummation
of the leading logarithmic (LL) corrections at the HERA kinematical
region ( $x\sim 10^{-3}$). 
The controversial aspect between their numerical analysis and 
the assertion by Bartels {\sl et al}. might be coming from the fact that
the resummed part of the \lq\lq coefficient function\rq\rq \ is
considered in Ref.~\cite{bartels} but not in Ref.~\cite{blvo}. 
Bl\"umlein and Vogt did not include the
resummed part of the coefficient function because this part
turns out to fall in the next-to-leading logarithmic (NLL) corrections
and depends on the factorization scheme adopted. 
It is also to be noted that the evolution, in general, strongly
depends on the input parton densities.
If one chooses a steep input function, the perturbative
contribution will be completely washed away since the structure function
is given as the convolution integral of the parturbative part and the
input density. So it will be interesting to see the sensitivity of the
results to the choice of the input densities.

In the present paper, we numerically 
reanalyse the flavor non-singlet part of
$g_{1}$ by taking into account the  $\ln x$ resummation.
We consider three different input densities: one is a flat
density corresponding to the naive Regge prediction and
others are steep ones in the small $x$ region.
The coefficient function can not
be included consistently at present since the anomalous
dimension has been calculated only at the LL order.
However we consider also the effects of the coefficient
function. The reason is because we could firstly clarify the
above controversial aspect and secondly get some
idea about the magnitude of the NLL order
corrections in the resummation approach.

This paper is organized as follows.
In section 2, we make a brief review on the resummation
of $\ln x$ series and present an explicit expression
for $g_{1}^{NS}$. 
In section 3, we show our numerical
results and discuss the effects of the NLL corrections.
The interpretation of the numerical results and
summary will be given in section 4. 


\section{Resummation of $\ln x$ terms } 

The flavor non-singlet part of the polarized structure
function $g^{NS}_{1}$
is given by the formula,
\be
 g^{NS}_{1}(Q^{2},x) = \frac{\langle e^{2} \rangle}{2}\int^{1}_{x}
   \frac{dy}{y} C^{NS}(\alpha_{s}(Q^{2}),x/y)
  \Delta q^{NS}(Q^{2},y) \ , \label{eqn:g1}
\ee
where $\Delta q^{NS}$ is the flavor non-singlet
combination of the polarized parton densities,
\[  \Delta q^{NS}(Q^{2},x) = \sum^{n_{f}}_{i=1}
  \frac{e_{i}^{2}-\langle e^{2} \rangle}{\langle e^{2} \rangle}
   (\Delta q_{i}(Q^{2},x) - \Delta \overline{q}_{i}(Q^{2},x)) \ ,
\]
and $C^{NS}$ is the coefficient function.
$n_{f}$ is the number of active flavors with electric charge 
$e_{i}$, $\langle e^{2} \rangle = \sum e_{i}^{2} / n_{f}$.
The perturbative evolution of the
parton density is controlled by the DGLAP
equation,
\be
 Q^{2} \frac{\partial}{\partial Q^{2}} \Delta q (Q^{2},x) 
  = 
 \int^{1}_{x}\frac{dy}{y} P (\alpha_{s}(Q^{2}),x/y)
      \Delta q (Q^{2},y) \ .\label{eqn:dgap}
\ee
In the above equation and in the following, we suppress
the superscript {\it NS} which means the flavor non-singlet part.
The coefficient function $C (\alpha_{s},y)$ and the
splitting function $P (\alpha_{s},y)$ are both 
calculable in the QCD perturbation theory.
When $x$ is finite, it may be enough to compute them
to the fixed-order of perturbation. 
In the small $x$ region, however, the fixed-order calculation
becomes questionable since there appear $\ln ^n x$ corrections
in the higher orders of the strong coupling constant $\alpha_s$.
If these $\ln ^n x$ terms
compensate the smallness of $\alpha_s$,
we must resum the perturbative series to the all orders to
get a reliable prediction.

To see what terms show up at small $x$, it will
be convenient to take the Mellin transform of Eq.(\ref{eqn:g1}).
\bean
  g_{1}(Q^{2},N) &\equiv&
   \int^{1}_{0}dx x^{N-1} g_{1}(Q^{2},x)\\
  &=& \frac{\langle e^{2} \rangle}{2}
     C (\alpha_{s}(Q^{2}),N) \Delta q (Q^{2},N)\ ,
\eean
where
\bean
  C (\alpha_{s}(Q^{2}),N) &\equiv& \int^{1}_{0}x^{N-1}
    C (\alpha_s (Q^{2}),x) \ ,\\
  \Delta q (Q^{2},N) &\equiv& \int^{1}_{0}x^{N-1}
    \Delta q (Q^{2},x)\ .
\eean
The DGLAP evolution equation Eq.(\ref{eqn:dgap}) becomes,
\be
  Q^{2} \frac{\partial}{\partial Q^{2}} \Delta q (Q^{2},N) 
   = - \gamma (\alpha_{s}(Q^{2}),N)
   \Delta q (Q^{2},N) \ .\label{eqn:rgeq} 
\ee
Here the anomalous dimension $\gamma$ is the moment of the
splitting function,
\[ \gamma (N,\alpha_{s}(Q^{2})) \equiv 
  - \int^{1}_{0}dx x^{N-1} P (\alpha_s (Q^{2}),x) \ .\]
Eq.(\ref{eqn:rgeq}) is easily solved to give, 
\[ \Delta q (Q^{2},N) = \Delta q (Q_{0}^{2},N)
    \exp \left( -\int^{\alpha_{s}(Q^{2})}_{\alpha_{s}(Q_{0}^{2})}
  \frac{d \alpha}{\beta(\alpha)}\gamma (\alpha,N) \right) \ ,\]
where $\beta$ is the beta function,
\[
   \beta (\alpha_{s}) = \frac{\partial \alpha_{s}}{\partial \ln Q^{2}}
       = \alpha_{s} \left[ -\beta_{0} \frac{\alpha_{s}}{4\pi}
   - \beta_{1}(\frac{\alpha_{s}}{4\pi})^{2} - \cdots  \right]
           \ .\]
The first two coefficients of the $\beta$ function are,
\[ \beta_{0} = \frac{11}{3} C_{A} - \frac{4}{3} T_{R}n_{f} \quad ,
   \quad \beta_{1} = \frac{34}{3} C_{A}^{2}
     - \frac{20}{3} C_{A} T_R n_f - 4 C_{F} T_{R} n_{f} \ ,\]
with
$C_{F}=(N_{c}^{2}-1)/2N_{c}$ and $C_{A}=N_{c}$ for
the SU($N_{c}$) color group and $T_R = 1/2$.

The coefficient function $C (\alpha_{s},N)$ and the
anomalous dimension $\gamma (\alpha_{s},N)$ may be expanded
in the powers of $\alpha_s$,
\bean
  C (\alpha_{s},N) &=& 1 + \sum_{k=1}^{\infty}
     c^{k}(N) \bar{\alpha}_{s}^{k} \ ,\\
  \gamma (\alpha_{s},N) &=& \sum_{k=1}^{\infty}
    \gamma^{k}(N) \bar{\alpha}_{s}^{k} \ .
\eean
where (and in the following) we use the abbreviation,
\[ \bar{\alpha}_{s} \equiv \frac{\alpha_s}{4\pi} \ .\]
The singular behaviors of the coefficient and splitting
functions as $x \to 0$ appear as the pole singularities at $N=0$
in the moment space.
The explicit next-to-leading order (NLO) calculations
of the coefficient function~\cite{koda} and the
anomalous dimension~\cite{van} in the $\overline{{\rm MS}}$ scheme
show a strong singularity at $N=0$,
\bea
 c^{1}(N) &=& 2 C_F \frac{1}{N^2} + {\cal O} \left( \frac{1}{N}
     \right)\ , \nonumber\\
 \gamma^{2}(N) &=& 4 (3 C_F^2 - 2 C_A C_F ) \frac{1}{N^3}
      + {\cal O} \left( \frac{1}{N^2} \right)\ , \label{NLO}
\eea
whereas the leading order anomalous dimension looks like,
\[ \gamma^{1}(N) = - 2 C_{F} \frac{1}{N} - C_{F}
     + {\cal O} (N) \ ,\]
at small $N$.
These strong singularities (double logarithmic corrections)
will persist to all orders of perturbative series.
Indeed, at the $k$-th loop, the anomalous dimension and the coefficient
function are expected to behave as,
\be
 \gamma^{k}(N) \sim N \left( \frac{1}{N^2} \right)^k \quad , \quad
   c^{k} (N) \sim \left( \frac{1}{N^2} \right)^k \ .\label{expect}
\ee
Since the singularities in $N^{-m}$ correspond to
the $\ln ^{m-1} \left( \frac{1}{x} \right)$ singularities,
our task is to resum these terms to all-orders in the perturbative
expansion.

Before discussing the resummed results, it may be worth mentioning the
difference between the polarized (unpolarized flavor non-singlet)
and the unpolarized flavor singlet structure functions~\cite{emr}.
Naively one expects that the anomalous dimension behaves like
$\gamma \sim \alpha_{s}^{k}/ N^{2k-1}$ at the $k$-th
loop\footnote{We follow the convention of Ref.~\cite{catani}
for the moment.}because there exist extra infrared and collinear
singularities.
In the case of the unpolarized flavor singlet structure functions,
however, many of them are canceled and the true behavior at the $k$-th
loop is $\gamma \sim (\alpha_{s}/ N )^{k}$.
These terms can be resummed by the BFKL equation.
On the other hand, above strong singularities survive in
the polarized structure function.
This fact suggests that the polarized structure function will
receive large perturbative corrections at small $x$.

The resummation of $\ln x$ singularities for the $g_1$ structure
function has been done in Refs.~\cite{kili}~\cite{bartels}.
The result for the \lq\lq parton (quark) \rq\rq\ target with
the fixed coupling constant\footnote{In the genuine LL approximation,
the strong coupling constant should be taken as a fixed parameter.}
reads,
\[
 g_{1}^{\rm parton} (x , Q^2 )= \frac{e_i^2}{2}
   \int_{ c-i \infty}^{c+i \infty} \frac{d N}{2 \pi i} x^{- N}
   \left( \frac{Q^2}{\mu^2} \right)^{f_0^- (N) / 8\pi^2}
   \frac{N}{N - f_0^- (N) / 8\pi^2} \ ,
\]
where $\mu$ is an arbitrary mass scale which regularizes the
infrared and/or mass singularities. From this expression we could
identify the resummed anomalous dimension $\hat{\gamma}$ and the
coefficient function $\hat{C}$ to be,
\bea
   \hat{\gamma}(\alpha_{s},N) 
    &\equiv& \lim_{N \to 0} \gamma(\alpha_{s},N) = -
   f^{-}_{0}(N)/8 \pi^{2} \ ,\label{resuma}\\
  \hat{C}(\alpha_{s},N) &\equiv& \lim_{N \to 0} C(\alpha_{s},N)
     = \frac{N}{N-f^{(-)}_{0}(N) / 8\pi^{2}} \ .\label{resumc}
\eea
Here $f_{0}^{-}$, which corresponds to the odd-signature
quark-quark scattering
amplitude in the color singlet channel, satisfies the equation,
\[ f_{0}^{-}(N) = 16 \pi^2 C_F \frac{\bar{\alpha}_s}{N}
   - 8 C_F \frac{\bar{\alpha}_s}{N^{2}} f_{8}^{+}(N)
    + \frac{1}{8\pi^{2}N}(f_{0}^{-} (N))^{2} \ .\]
The even-signature quark-quark 
scattering amplitude $f_{8}^{+}$ with the color octet quantum
number in the t-channel is the solution of the
equation,
\[ f_{8}^{+}(N) = - 8 \pi^2 \frac{1}{N_c} \frac{\bar{\alpha}_s}{N}
  + 2 N_C \frac{\bar{\alpha}_s}{N} \frac{d}{dN}f_{8}^{+}(N) 
  + \frac{1}{8\pi^{2}N}(f_{8}^{+} (N))^{2} \ ,\]
and given by,
\[ f^{+}_{8} (N) = 16 \pi^2 N_{c} \bar{\alpha}_s \frac{d}{dN}
   \ln(e^{z^{2}/4}D_{-1/2N_{c}^{2}}(z)) \qquad {\rm with}
  \qquad z = \frac{N}{\sqrt{2 N_{c} \bar{\alpha}_s}}\ .\]
$ D_{p}(z)$ is the
parabolic cylinder function~\cite{Grad}.
Finally we reach,
\[   f^{-}_{0}(N) = 4 \pi^{2}N
  \left( 1-\sqrt{1- 8 C_F \frac{\bar{\alpha}_s}{N^{2}}
  \left[ 1-\frac{1}{2\pi^{2}N}f^{(+)}_{8}(N) \right]} \right) \ .\]

Now it will be instructive to re-expand Eqs.(\ref{resuma},\ref{resumc})
in terms of $\alpha_s$ to see whether these formulae sum up the
most singular terms of the perturbative series.
The expressions expanded up to ${\cal O} (\alpha_s^5 )$ read,
\bea
  \hat{\gamma} &=& - N \left[
     2 C_{F} \left( \frac{\bar{\alpha}_{s}}{N^{2}} \right)
   + 4 C_{F} \left(C_{F}+\frac{2}{N_{c}} \right)
     \left( \frac{\bar{\alpha}_{s}}{N^{2}} \right)^{2} \right.
            \nonumber \\
   & & \ \ + \, 16 C_F \left( C_{F}^2 + 2 \frac{C_{F}}{N_{c}}
       - \frac{1}{2N_{c}^{2}} - 1 \right)
    \left( \frac{\bar{\alpha}_{s}}{N^{2}} \right)^{3} \nonumber\\
   & & \ \ + \, \left. 16 C_{F} \left( 5 C_F^3 +
     12 \frac{C_F^2}{N_c} + 2 \frac{C_F}{N_c^2} - 4 C_F
      + \frac{1}{N_c^3} + \frac{5}{N_c} + 6 N_c \right)
        \left( \frac{\bar{\alpha}_{s}}{N^{2}} \right)^{4} \right]
     \nonumber\\
   & & \ \ + \, \cdots \nonumber\\
   &=& N \sum^{\infty}_{k=1} \hat{\gamma}^{k}
   \left( \frac{\bar{\alpha}_{s}}{N^{2}} \right)^{k} \ ,\label{ptg}\\
 \hat{C} &=& 1 + 2C_{F} \left( \frac{\bar{\alpha}_{s}}{N^{2}} \right)
     + 8 C_F \left( C_{F} + \frac{1}{N_{c}} \right)
   \left( \frac{\bar{\alpha}_{s}}{N^{2}} \right)^{2} \nonumber \\
   & & \ \ + \, 8 C_F \left( 5 C_{F}^{2} + 8 \frac{C_F}{N_c}
        - \frac{1}{N_c^2} - 2 \right)
     \left( \frac{\bar{\alpha}_{s}}{N^{2}} \right)^{3} \nonumber \\
   & & \ \ + \, 32 C_F \left( 7 C_{F}^{3} + 15 \frac{C_F^2}{N_{c}}
      + 2 \frac{C_{F}}{N_{c}^{2}} - 4 C_F
      + \frac{1}{2 N_c^3} + \frac{5}{2} \frac{1}{N_c} + 3 N_c \right)
     \left( \frac{\bar{\alpha}_{s}}{N^{2}} \right)^{4}  
     \nonumber\\
   & & \ \ + \, \cdots \nonumber\\
   &=& \sum^{\infty}_{k=0} \hat{c}^{k}
   \left( \frac{\bar{\alpha}_{s}}{N^{2}} \right)^{k} \ . \label{ptc}
\eea
These results coincide with the
previous expectation of Eq.(\ref{expect}). Furthermore,
noting the relation,
\[ 2 C_{A}- 3 C_{F} = \frac{2}{N_{c}} + C_{F} \ ,\]
which holds in SU($N_{c}$), we can see that the resummed expressions
Eqs.(\ref{resuma},\ref{resumc}) reproduce
the known NLO results Eqs.(\ref{NLO}) in the $\overline{\rm MS}$ scheme.
Therefore, it is quite plausible that
Eqs.(\ref{resuma},\ref{resumc}) correctly sum up
the \lq\lq leading\rq\rq\ singularities to all orders.

Here a comment is in order concerning the scheme dependence.
It is well-known that the anomalous dimension and the coefficient
function individually depend on the factorization scheme and
only an appropriate
combination of them becomes scheme independent.
When one considers the higher order corrections in the perturbation
theory, therefore, one must specify the scheme adopted. 
Unfortunately we do not have by now any appropriate factorization
theorems to the problem discussed in this paper.
This means that we must be careful when considering
the resummed quantities.
In particular, the resummed \lq\lq coefficient function\rq\rq\ does
not have any physical meaning until the scheme dependent part
of the anomalous dimension is calculated in the same scheme.
To clarify this issue, it is convenient to write
the above results in the form which corresponds
to the so-called DIS scheme~\cite{aem}.
The DIS scheme is defined so
that the naive parton model relation is true to all orders
in perturbation theory.
The polarized parton densities become physical observables in
this scheme.
The parton densities and anomalous dimension in the DIS scheme
are obtained by making the transformations,
\[ \Delta q
 \rightarrow  \Delta q^{DIS} \equiv C \Delta q \ ,\]
\[ \gamma^{DIS} \equiv C \gamma C^{-1}
      - \beta(\alpha_{s})
    \frac{\partial}{\partial \alpha_{s}} \ln C \ .\]
Using the resummed
$\hat{\gamma}$ and $\hat{C}$ Eqs.(\ref{ptg},\ref{ptc}),
we get the resummed part of the anomalous dimension in the DIS scheme,
\be
 \hat{\gamma}^{DIS} = N \sum^{\infty}_{k=1}\hat{\gamma}^{k}
      \left( \frac{\bar{\alpha}_{s}}{N^{2}} \right)^{k}
  + \beta_{0} N^2 \sum^{\infty}_{k=2}
    \hat{d}^{k}
    \left( \frac{\bar{\alpha}_{s}}{N^{2}} \right)^{k} 
  + {\cal O} \left( N^3 
      \left( \frac{\bar{\alpha}_{s}}{N^{2}} \right)^{k} \right) \ ,
     \label{resumdis}
\ee
where the second terms come from the resummed coefficient function
and $\hat{d}^k$ are numerical numbers independent of $N$.
The above equation tells us that the resummed coefficient function
belongs to the NLL order\footnote{This fact implies the LL
resummed anomalous dimension $\hat{\gamma}$ being scheme independent.}
corrections in the context of the resummation approach.
Then, one must include the NLL order anomalous
dimension which has not yet been available
to see the effects of the coefficient function.
This is the reason why the authors in Ref.~\cite{blvo} throw away
the coefficient function.


\section{Numerical Analysis}

Numerical analysis of the spin structure function $g_{1}^{NS}$  
in the small $x$ region was done in the context of
the small $x$ resummation approach in Ref.~\cite{blvo}.
They obtained the result that the small
$x$ resummation effect is not significant despite of
a naive expectation discussed in Ref.~\cite{bartels}.
In this section, we numerically reanalyze the behavior of 
$g_{1}^{NS}$ structure function to show how the final results
are sensitive to the choice of the input parton densities.
In conjunction with the claim in Ref.~\cite{bartels},
we also consider the effects from the resummed coefficient function.
As already discussed in section 2, we can not include the coefficient
function in a theoretically consistent way. However we believe
that the inclusion of the coefficient function could shed some light
on the size of the NLL corrections in the resummation
approach.

At first, we explain our method to estimate the $g_{1}^{NS}$
structure function numerically.
Our starting point is the expression,
\be
 g_{1}^{NS}(Q^{2},x) = \int_{ c-i \infty}^{c+i \infty} 
    \frac{d N}{2 \pi i} x^{- N} \exp
\left(-\int_{ \alpha_{s}(Q_{0}^{2})}^{\alpha_{s}(Q^{2})}
       \frac{d\alpha_{s}}{\beta} \gamma^{DIS} \right)
           g_{1}^{NS}(Q_{0}^{2},N) \ .
\label{eqn:g1-DIS}
\ee
The anomalous dimension $\gamma^{DIS}$ which includes the resummation
of ${\ln}^n x$ terms is organized as follows,
\be
\gamma^{DIS}(N) = \bar{\alpha}_{s} \gamma^{1}(N)
   + \bar{\alpha}_{s}^{2} \gamma^{2}(N) + K (N,\alpha_{s})
  - \beta \frac{\partial}{\partial \alpha_{s}}
   \ln \left( 1 + \bar{\alpha}_{s} c^{1} + H(N,\alpha_{s}) \right) \ ,
\label{eqn:anoma}
\ee
where $\gamma^{1,2}$ and $c^{1}$ are respectively the usual anomalous
dimension and coefficient function at the one and two-loop fixed order
perturbation theory.
$K(N,\alpha_{s})$ ($H(N,\alpha_{s})$) is the resummed anomalous
dimension Eq.(\ref{ptg}) (Eq.(\ref{ptc})) with $k = 1,2$ ($k = 0,1$)
terms being subtracted because those terms have already been
included in the usual anomalous
dimension and coefficient function.
\bean
 K(N,\alpha_{s}) &\equiv&
        \hat{\gamma}(N) - \hat{\gamma}^1 \frac{\bar{\alpha}_{s}}{N}
    - \hat{\gamma}^2 \frac{\bar{\alpha}_{s}^{2}}{N^{3}} \ ,\\
 H(N,\alpha_{s}) &\equiv&
        \hat{C}(N) - 1 - \hat{c}^1 \frac{\bar{\alpha}_{s}}{N^{2}}\ .
\eean

It should be noted here that the anomalous dimension at $N=1$ plays
a special role for the non-singlet $g_1$ structure function.
In a language of the operator product expansion,
$\gamma (N=1)$ is the anomalous dimension of the (non-singlet) axial
vector current.
Since the (non-singlet) axial vector current is conserved,
the corresponding anomalous dimension should vanish.
The perturbation theory guarantees this symmetry order by order
in the $\alpha_{s}$ expansion. However, the
resummation of the leading singularities in $N$ does not respect
this symmetry. Therefore, we need to restore this symmetry 
\lq\lq by hand\rq\rq . In this paper, we multiply
$K(N,\alpha_{s})$ by $(1-N)$~\cite{Ellis},
\[  K(N,\alpha_{s}) \to K(N,\alpha_{s}) (1-N) \ .\]
Of course, this is not a unique prescription and one can choose other
procedure\footnote{Our final conclusion remains the same qualitatively
if we choose other prescription.}
which satisfies the condition
of $\lim_{N \rightarrow 1}K(N,\alpha_{s}) = 0$.

Now let us explain how to perform
the Mellin inversion Eq.(\ref{eqn:g1-DIS})
which is the integral in the complex $N$-plane.
At first, we must know the Mellin transform of the input function
$g_{1}(Q_0^{2},N)$. It is easy to obtain an analytical
form for it in the complex $N$-plane since we assume a simple function
(see below) for the input density.
Next we need an analytically continued expression of
the anomalous dimension $\gamma^{DIS}$ in the complex $N$-plane.
For the $g_{1}$ structure function, only odd moments are defined.
So we replace $(-1)^{N}$ by $(-1)$ in the expression of the 
anomalous dimension obtained in Ref.~\cite{van}.
The integration contour in the Mellin inversion should be on the right
of the rightmost singularity of the integrand.
The contour integration along the imaginary axis
from $c-i \infty$ to $c + i \infty$
is numerically inconvenient due to the slow convergence of the integral
in the large $|N|$ region.
To get rid of this problem, we deformed the contour to the
line which have an angle $\phi$ ($\phi > \pi / 2$) from the real
$N$ axis. By this change of the contour, we have a damping factor
$\exp (|N| \ln (1/x) \cos\phi )$ which 
strongly suppresses the contribution from the large $|N|$ region.
In the integration along this new contour, we will be able to cut the
large $|N|$ region. Finally we have checked the stability of results
by changing the contour parameter.
One can find the details of this technique in Ref.~\cite{Reya}.

We choose the starting value of the evolution to be
$Q_{0}^{2} = 4 GeV^{2}$. We calculate the $Q^2$ evolution for
three types of the input densities A, B and C:
A is a function which is flat at small $x$ ($x^{\alpha},\alpha \sim 0$),
B is slightly steep\footnote{This choice is essentially the same as one
in Ref.~\cite{blvo}.} ($\alpha \sim - 0.2$) and
C rises more steeply ($\alpha \sim - 0.7$).
The explicit parametrization used in this paper is~\cite{bfr},
\[ \Delta q (Q_0^{2},x) = N (\alpha , \beta , a )
       \eta x^{\alpha} (1 - x)^{\beta} ( 1 + a x)\ ,\]
where $N$ is a normalization factor such that
$\int dx N x^{\alpha} (1 - x)^{\beta} ( 1 + a x) = 1$
and $\eta = \frac{1}{6} g_A $ ($g_A = 1.26 $) in
accordance with the  Bjorken sum rule.
A, B and C correspond to the following values of parameters,
\bean
    A &:& \alpha = + 0.0 \  , \ \beta = 3.09 \  , \ a = 2.23 \ ,\\
    B &:& \alpha = - 0.2 \  , \ \beta = 3.15 \  , \ a = 2.72 \ ,\\
    C &:& \alpha = - 0.5 \  , \ \beta = 2.41 \  , \ a = 0.02 \ .
\eean
In our analysis we put the flavor number 
$n_{f} = 4 $ and $\Lambda_{QCD}=0.23 GeV$.

First we estimate the case which includes only the LL correction
$\hat{\gamma}$.
The evolution kernel in this case is obtained by dropping   
$H(N,\alpha_{s})$ in Eq.(\ref{eqn:anoma})\footnote{We take into account
the fixed-order NLO corrections exactly.}.
This is a consistent approximation in the resummation approach.
Fig.1a (1b, 1c) shows the results (dashed curves) after evolving
to $Q^{2} = 10, 10^2 ,10^4 GeV^{2}$ from the A (B, C) input density
(dot-dashed line). The solid curves are the predictions of
the NLO-DGLAP evolution. 
These results show a tiny enhancement compared with the NLO-DGLAP
analysis and are consistent with those in Ref.~\cite{blvo}\footnote{We
have also calculated $g_1$ with the input function used
by Bl\"umlein and Vogt and could reproduce their results.}.
In the case of C, we can not discriminate
a difference between the LL and DGLAP results.
The enhancement is, as expected, bigger when the input density is
flatter. However any significant differences are not seen between the
results from different input densities.

Next, we include the NLL corrections coming from the resummed
\lq\lq coefficient function\rq\rq .
We show the results in Fig.2 by the dashed curves.
(Other curves are the same as in Fig.1.)
The results are rather surprising. The inclusion of
the coefficient function leads to a strong suppression
on the evolution of the structure function at small $x$.
Since the effects from the coefficient function fall in the NLL
level, the LL terms are
expected to (should) dominate in the small $x$. However our results
imply that the LL approximation
is not sensible in the small $x$ region we are interested in.
As the resummed coefficient function is only a part of the
NLL correction, we can not present a definite conclusion on the
(full) NLL correction. But it is obvious that the NLL correction
is very important at the experimentally accessible region of $x$.
In the following section, we explain why the coefficient function
leads to such suppression.


\section{Discussion and Summary}

In the previous section, we have shown that although the LL resummed
effect is very small at the experimentally accessible region of $x$,
a part of the NLL resummed contribution from the coefficient function
drastically changes the predictions.

To understand these numerical results, it will be helpful to remember the
perturbative
expansion of the resummed anomalous dimension and coefficient function
Eqs.(\ref{ptg},\ref{ptc}).
By using the explicit values $N_C = 3 , C_F = 4/3$, we obtain
for the anomalous dimension in the DIS scheme
Eq.(\ref{resumdis}),
\bea
 \hat{\gamma}^{DIS} &=& N \left[ - 0.212
       \left( \frac{\alpha_{s}}{N^{2}} \right) \right. \nonumber\\
    & & \qquad - \, 
       0.068 \left. \left( \frac{\alpha_{s}}{N^{2}} \right)^{2} -
       0.017 \left( \frac{\alpha_{s}}{N^{2}} \right)^{3} -
       0.029 \left( \frac{\alpha_{s}}{N^{2}} \right)^{4} 
           + \cdots \right] \nonumber\\
    &+&  \, N^2 \left[
       0.141 \left( \frac{\alpha_{s}}{N^{2}} \right)^{2} +
       0.119 \left( \frac{\alpha_{s}}{N^{2}} \right)^{3} +
       0.069 \left( \frac{\alpha_{s}}{N^{2}} \right)^{4} 
           + \cdots \right] \label{numbergdis}\\ 
    &+&  \,\,\,  \cdots \ \ .\nonumber
\eea
Here note that: (1) the perturbative coefficients of the LL terms
(the first part of Eq.(\ref{numbergdis})) are negative
and those of the higher orders are rather small number.
This implies that the LL corrections push up
the structure function compared to the fixed-order DGLAP
evolution, but the deviations are expected to be small. 
(2) the perturbative ones from the NLL terms (the second part of
Eq.(\ref{numbergdis})), however, are positive and
somehow large compared with those of the LL terms.
This positivity of the NLL terms has the effect of
decreasing the structure function. This fact that
the coefficients with both sign appear in the anomalous dimension
should be contrasted with
the case of the unpolarized structure function~\cite{catani2}.

Now it might be also helpful to {\sl assume} that the saddle-point dominates
the Mellin inversion Eq.(\ref{eqn:g1-DIS}). We have numerically estimated
the approximate position of the
saddle-point and found that the saddle-point stays around
$N_{\rm SP} \sim 0.31$ in the region of $x \sim 10^{-5}$ to $10^{-2}$.
(Of course the precise value of the saddle-point depends on
$x, Q_0^2$ and $Q^2$.) By looking at the explicit values of the coefficients in
Eq.(\ref{numbergdis}), the position of the saddle-point seems to suggest that
the NLL terms can not be neglected. Since the coefficients from the higher
order terms are not so large numerically, it is also
expected that the terms  which
lead to sizable effects on the evolution may be only first few terms in the
perturbative series in the region of $x$ we are interested in.
We have checked that the inclusion of the first few terms in
Eq.(\ref{numbergdis}) already reproduces the results of section 3.
Fig.3a (3b) shows the numerical results of the
contribution from each terms of the NLL corrections in
Eq.(\ref{numbergdis}) at $Q^{2} = 10^2 GeV^{2}$ with the A (B)
type input density.
The solid (dot-dashed) line corresponds to the NLL (LL) result.
The long-dashed, dashed and dotted lines correspond
respectively to the case in which
the terms up to the order $\alpha_{s}^2$, $\alpha_{s}^3$, $\alpha_{s}^4$,
are kept in the NLL contributions. One can see that the
dotted line already coincides with the full NLL (solid) line.
These considerations could help us to understand why the NLL corrections
turns out to give large effects on the evolution of the $g_1$
structure function.


The final discussion concerns the convergence issue of the perturbative
series. As discussed in Refs.~\cite{bf}~\cite{frt}, one must be careful
when applying the perturbative approach to small $x$ evolution.
The integrand in Eq.(\ref{eqn:g1-DIS}) has a singularity in the
moment space. This (rightmost) singularity 
is equal to that of $f_0^- (N)$. The numerical value $N_0$ of
the singularity position is $N_0 \sim 0.304$. This means that
the $N$ can not become so small. On the other hand, the approximation
scheme in the resummation approach is sensible only for small $N$.
This apparent contradiction will be solved by analyzing the evolution
in $x$ space~\cite{bf}. By explicitly solving the evolution
in $x$ space, it has been pointed out~\cite{frt}
that the saddle-point method is not a good approximation in the case
of the unpolarized structure function. Although we have not
used the saddle-point approximation to solve the evolution,
the previous explanation relying on this method can be misleading.
So according to Refs.~\cite{bf}~\cite{frt}, we have tried to solve
the evolution in $x$ space with first several terms
of the perturbative expansion being
kept and what we found is that the conclusion does not change.
The numerical results are essentially the same as Fig.3.

In summary, we have performed numerical studies
for the flavor non-singlet $g_{1}$ structure function at small $x$
by incorporating the all-order resummed anomalous dimension
and a part of the NLL corrections from the resummed coefficient
function.
Our results show that the resummed coefficient function has an effect
which suppresses the structure function at small $x$.
Including only the resummed coefficient
part is not theoretically consistent, and so one should take
into account also the anomalous dimension at the NLL level.
However, our results suggest that the LL analysis is unstable,
in the sense that a large suppression effect comes from the resummed
coefficient function which should be NLL correction.
We have explained why the inclusion of a part of the NLL corrections
leads to such unexpected results.
We need a full NLL analysis to make a definite conclusion
on whether the all-order resummation approach
predicts a rise of the flavor non-singlet $g_1$ structure
function in the experimentally accessible small $x$ region.

\vspace{1cm}
\noindent
{\large\bf Acknowledgements}

\bigskip
\noindent
J.K. would like to thank the
Institute for Nuclear Theory at the University
of Washington for its hospitality and the Department of Energy
for partial support at the beginning of this work.
He is also very grateful to Professors W. Bardeen
and Stephen Parke for the hospitality extended to him at FNAL where
part of this work was done.
Y.K. would like to thank K.Ochi for careful looking
over the FORTRAN Program. H.T. thanks M.Hirata, K.Tanaka and
T.Onogi for discussions.
We also would like to thank Tsuneo Uematsu for valuable comments
and helpful suggestions. J.K. was supported in part by the Monbusho
Grant-in-Aid Scientific Research No. A (1) 08304024.


\newpage
\noindent
{\large\bf Figure Captions}

\medskip
\noindent
Fig. 1

\noindent
The LL evolution as compared to the DGLAP results with the flat
input A (Fig. 1a) and steep ones B (Fig. 1b) and C (Fig. 1c).

\medskip
\noindent
Fig. 2

\noindent
The NLL evolution as compared to the DGLAP results with the
flat input A (Fig. 2a) and steep ones B (Fig. 2b) and C (Fig. 2c).

\medskip
\noindent
Fig. 3

\noindent
Contributions from the fixed order terms in the NLL
resummation with the flat input A (Fig. 3a) and steep one B
(Fig. 3b).


\newpage
\baselineskip 18pt

\begin{thebibliography}{99}
%
\bibitem{intro}
   See {\sl e.g.} J. Kwieci\'nski, hep-ph/9607221;\\
    B.~R.~Webber, {\sl Preprint} Cavendish-HEP-96/2 hep-ph/9607441;\\
    J. Bl\"umlein, S. Riemersma and A. Vogt, {\sl Preprint}
    DESY 96-131 / WUE-ITP-96-016 hep-ph/9608470; hep-ph/9610427
    and references therein.

\bibitem{bfkl}
   L. N. Lipatov, {\sl Sov. J. Nucl. Phys.} {\bf 23} (1976) 338;\\
   E. A. Kuraev, L. N. Lipatov and V. S. Fadin, {\sl Sov. Phys. JETP}
   {\bf 45} (1977) 199;\\
   Ya. Balitskii and L. N. Lipatov, {\sl Sov. J. Nucl. Phys.} {\bf 28}
   (1978) 822.

\bibitem{catani}
   S. Catani and F. Hautmann, {\sl Nucl. Phys.} {\bf B427} (1994) 475
   and references therein.

\bibitem{smc}
   J.~Ashman {\it et al.} {\sl Phys. Lett.} {\bf B206} (1988) 364;\\
   V.~W.~Hughes {\it et al.} {\sl Phys. Lett.} {\bf B212} (1988) 511;\\
   B.~Adeva {\it et al.} {\sl Phys. Lett.} {\bf B302} (1993) 553;
   {\bf B320} (1994) 400; \\
   D.~Adams {\it et al.}, {\sl Phys. Lett.} {\bf B329} (1994) 399;
   {\bf B336} (1994) 125; \\
   P.~L.~Anthony {\it et al.} {\sl Phys. Rev. Lett.} {\bf 71} (1993) 959;\\
   K.~Abe {\it et al.} {\sl Phys. Rev. Lett.} {\bf 74} (1995) 346;
   {\sl ibid.} {\bf 75} (1995) 25;
   {\sl ibid.} {\bf 76} (1996) 587.

\bibitem{heimann}
  R.~L.~Heimann, {\sl Nucl. Phys.} {\bf B64} (1973) 429.

\bibitem{dglap}
  G.~Altarelli, {\sl Phys. Rep.} {\bf 81} (1982) 1
     and references therein.

\bibitem{bfr}
   R. D. Ball, S. Forte and G. Ridolfi, {\sl Nucl. Phys.} {\bf B444}
   (1995) 287; {\sl Phys. Lett.} {\bf B378} (1996) 255.

\bibitem{grsv}
   M. Gl\"uck, E. Reya and W. Vogelsang, {\sl Phys. Lett.}
   {\bf B359} (1995) 201;\\
   M. Gl\"uck, E. Reya, M. Stratmann and W. Vogelsang, {\sl
   Phys. Rev.} {\bf D53} (1996) 4775.

\bibitem{kili}
   R. Kirschner and L. N. Lipatov, {\sl Nucl. Phys.} {\bf B213}
         (1983) 122.

\bibitem{bartels}
   J. Bartels, B. I. Ermolaev and M. G. Ryskin,
   {\sl Z. Phys.} {\bf C70} (1996) 273; hep-ph/9603204

\bibitem{blvo}
   J. Bl\"umlein and A. Vogt ,{\sl Phys. Lett.} {\bf B370}
   (1996) 149 ; {\sl Acta.Phys.Polonica} {\bf B27} (1996) 1309; \\
   J. Bl\"umlein, S. Riemersma and A. Vogt, hep-ph/9608470

\bibitem{koda}
   J. Kodaira, S. Matsuda, T. Muta, K. Sasaki and T. Uematsu,
      {\sl Phys. Rev.} {\bf D20} (1979) 627; \\
   J. Kodaira, S. Matsuda, K. Sasaki and T. Uematsu,
      {\sl Nucl. Phys.} {\bf B159} (1979) 99.

\bibitem{van}
    E. G. Floratos, D. A. Ross and C. T. Sachrajda, {\sl Nucl. Phys.}
    {\bf B129} (1977) 66; (E): {\bf B139} (1978) 545;
        {\bf B152} (1979) 493;\\
    A. Gonzalez-Arroyo, C. Lopez and F. J. Yndurain,
    {\sl Nucl. Phys.} {\bf B153} (1979) 161;\\
    A. Gonzalez-Arroyo and C. Lopez, {\sl Nucl. Phys.} {\bf B166}
   (1980) 429;\\
    E. G. Floratos, C. Kounnas and R. Lacaze, {\sl Nucl. Phys.} {\bf B192}
    (1981) 417; {\sl Phys. Lett.} {\bf B98} (1981) 89;\\
    G. Curci, W. Furmanski and R. Petronzio, {\sl Nucl. Phys.}
    {\bf B175} (1980) 27;\\
    W. Furmanski and R. Petronzio, {\sl Phys. Lett.} {\bf B97} (1980)
    437 ; {\sl Z. Phys.} {\bf C11} (1982) 293 ;\\
    R. Mertig and W. L. van Neerven, {\sl Z. Phys.} {\bf C70}
     (1996) 637.

\bibitem{emr}
   B. I. Ermolaev, S. I. Manayenkov and M. G. Ryskin,
   {\sl Z. Phys.} {\bf C69} (1996) 259.

\bibitem{Grad}
   I. S. Gradshtein and I. M. Ryzhik, {\sl Table of Integrals,
   Series and Products} ({\bf DVW, Berlin}) Section 9.24-9.25.

\bibitem{aem}
   G. Altarelli, R. K. Ellis and G. Martinelli, {\sl Nucl. Phys.}
   {\bf B157} (1979) 461.

\bibitem{Ellis}
    K. Ellis, F. Hautmann and B. Webber, {\sl Phys. Lett.}
      {\bf B348} (1995) 582.

\bibitem{Reya}
   M. Gl\"uck, E. Reya and A. Vogt, {\sl Z. Phys.}
      {\bf C48} (1990) 471; \\
   D. Graudenz, M. Hampel, A. Vogt and Ch. Berger, {\sl Z. Phys.}
      {\bf C70} (1996) 70.

\bibitem{catani2}
   S. Catani, {\sl Z. Phys.} {\bf C70} (1996) 263.

\bibitem{bf}
   R. D. Ball and S. Forte, {\sl Phys. Lett.} {\bf B351} (1995) 313.

\bibitem{frt}
   J. R. Forshaw, R. G. Roberts and R. S. Thorne,
   {\sl Phys. Lett.} {\bf B356} (1995) 79.

\end{thebibliography}
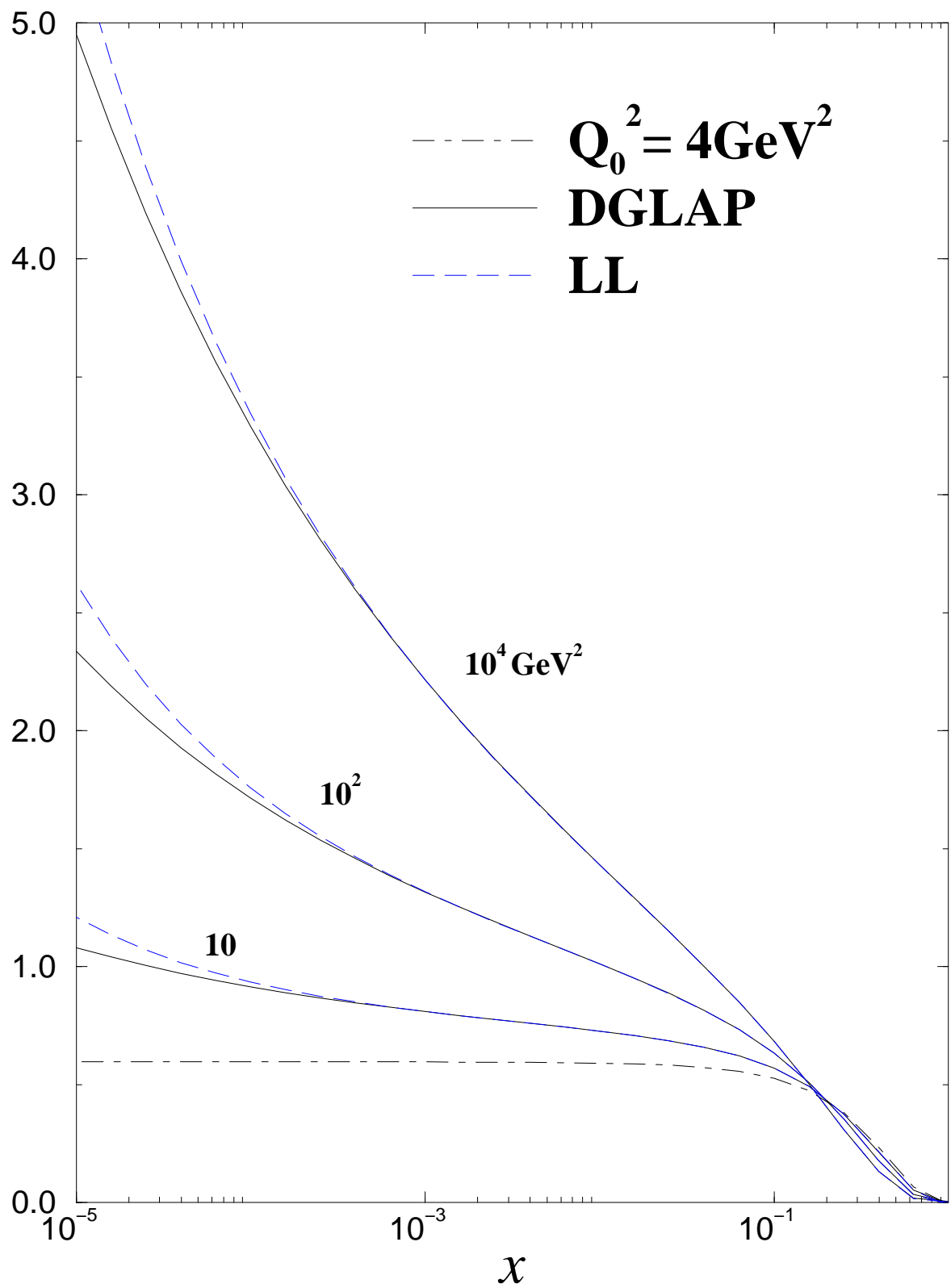
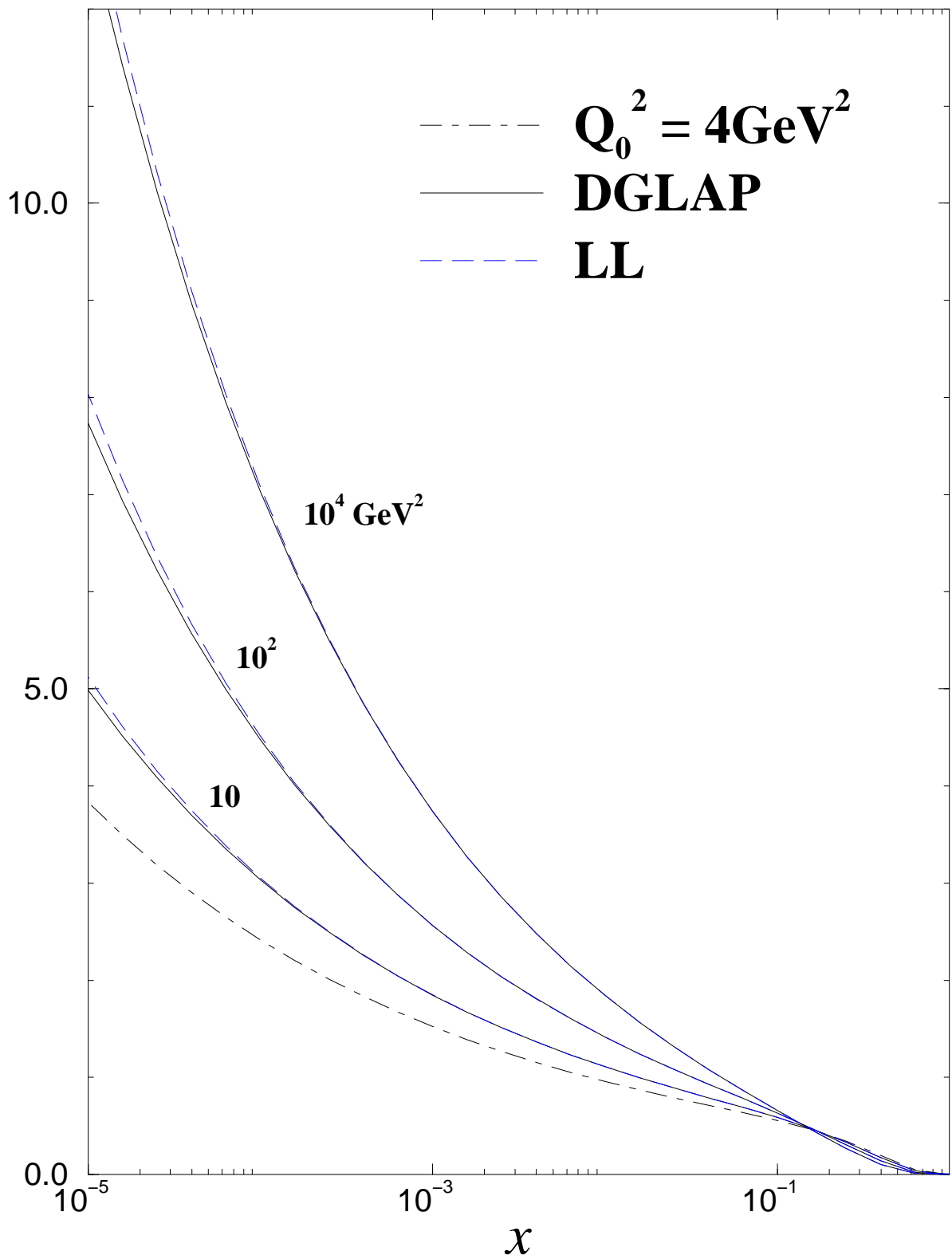
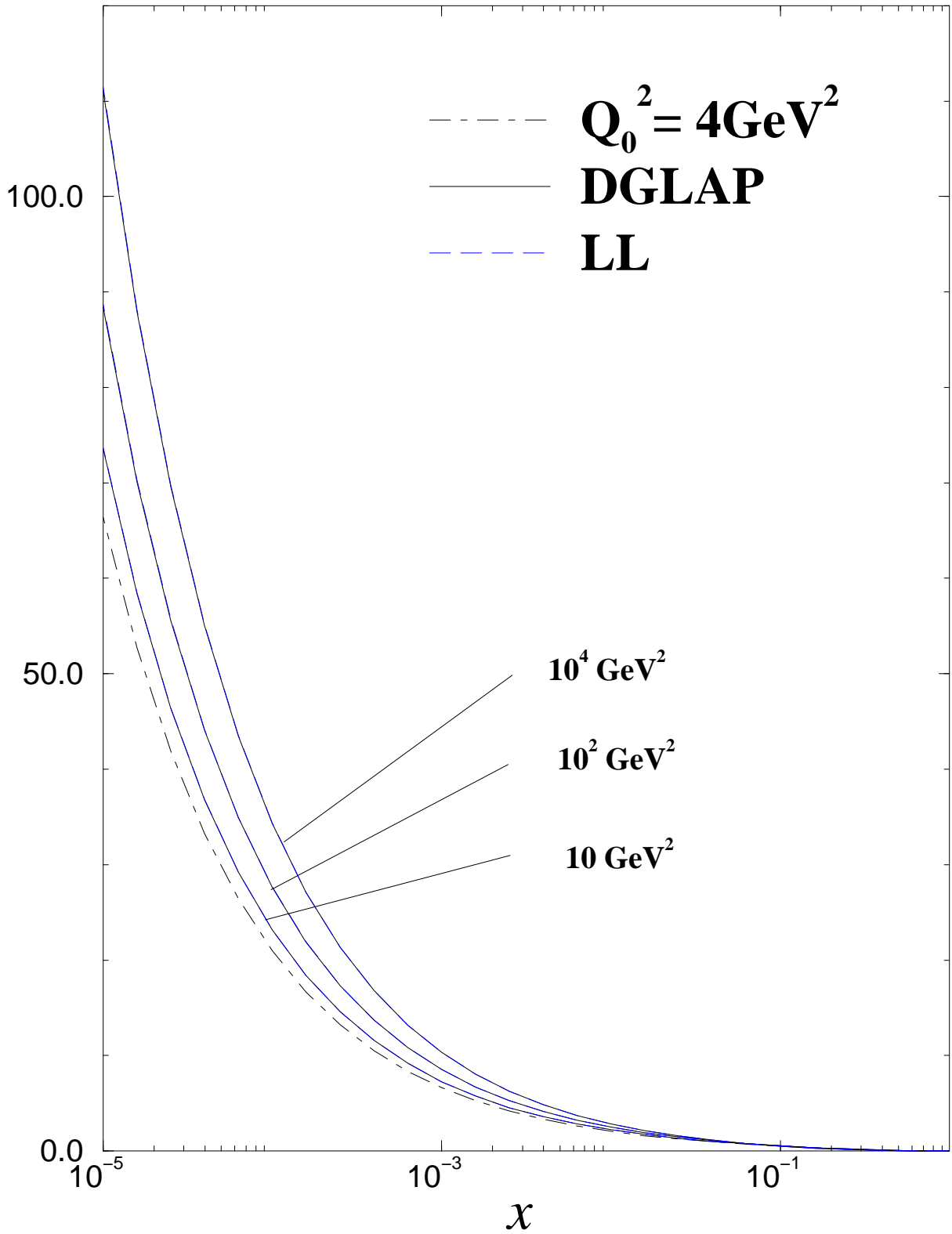
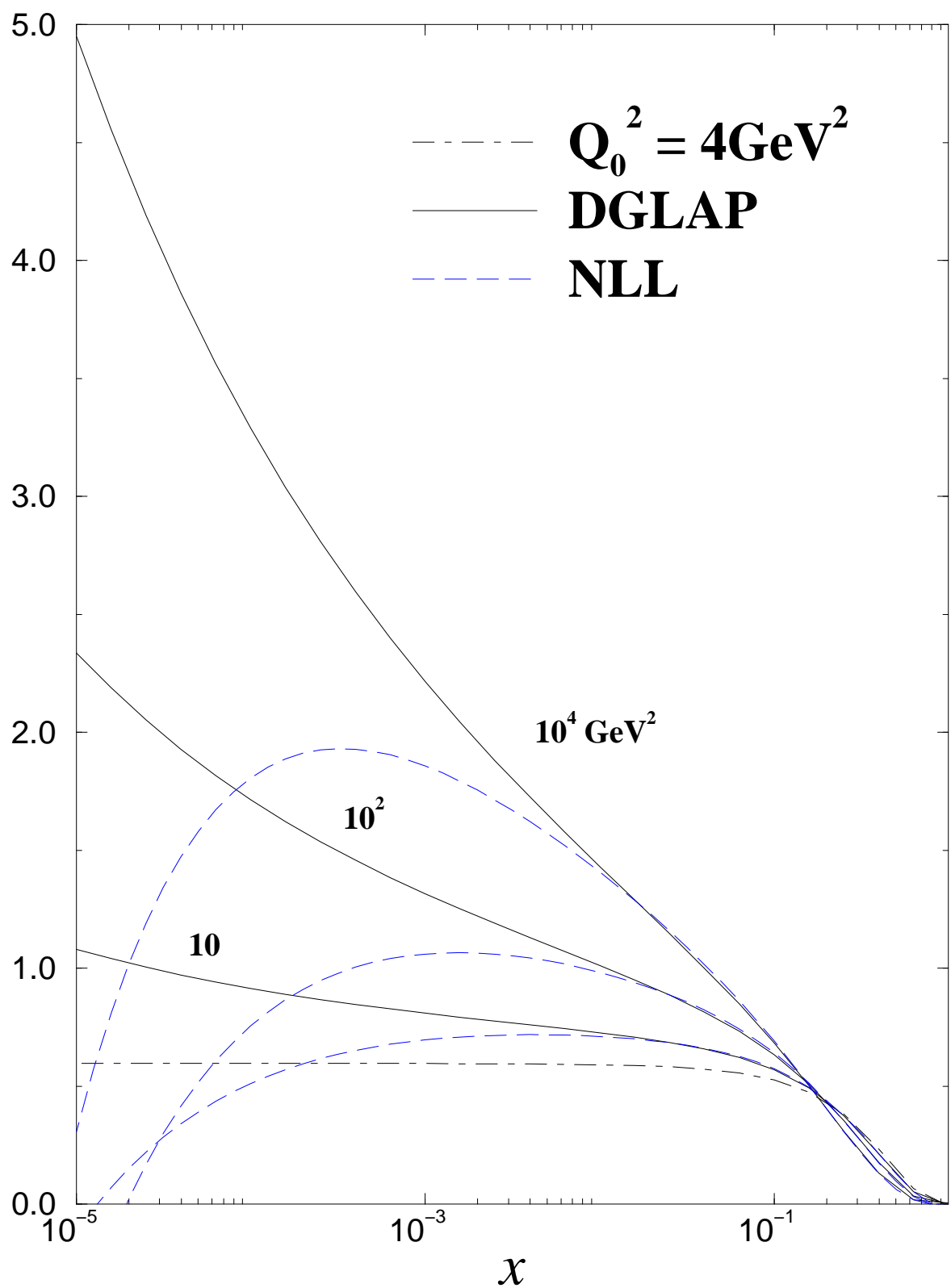
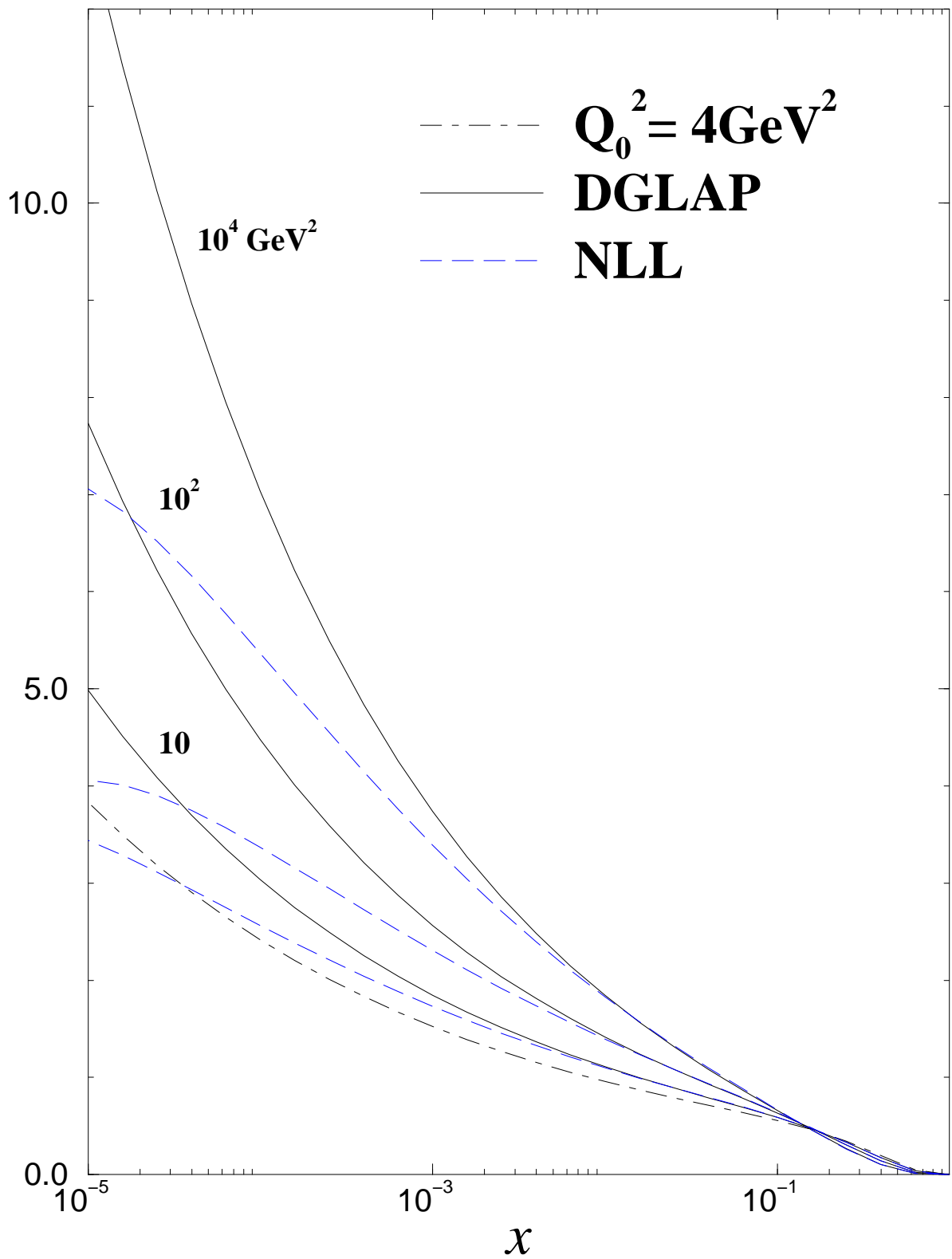
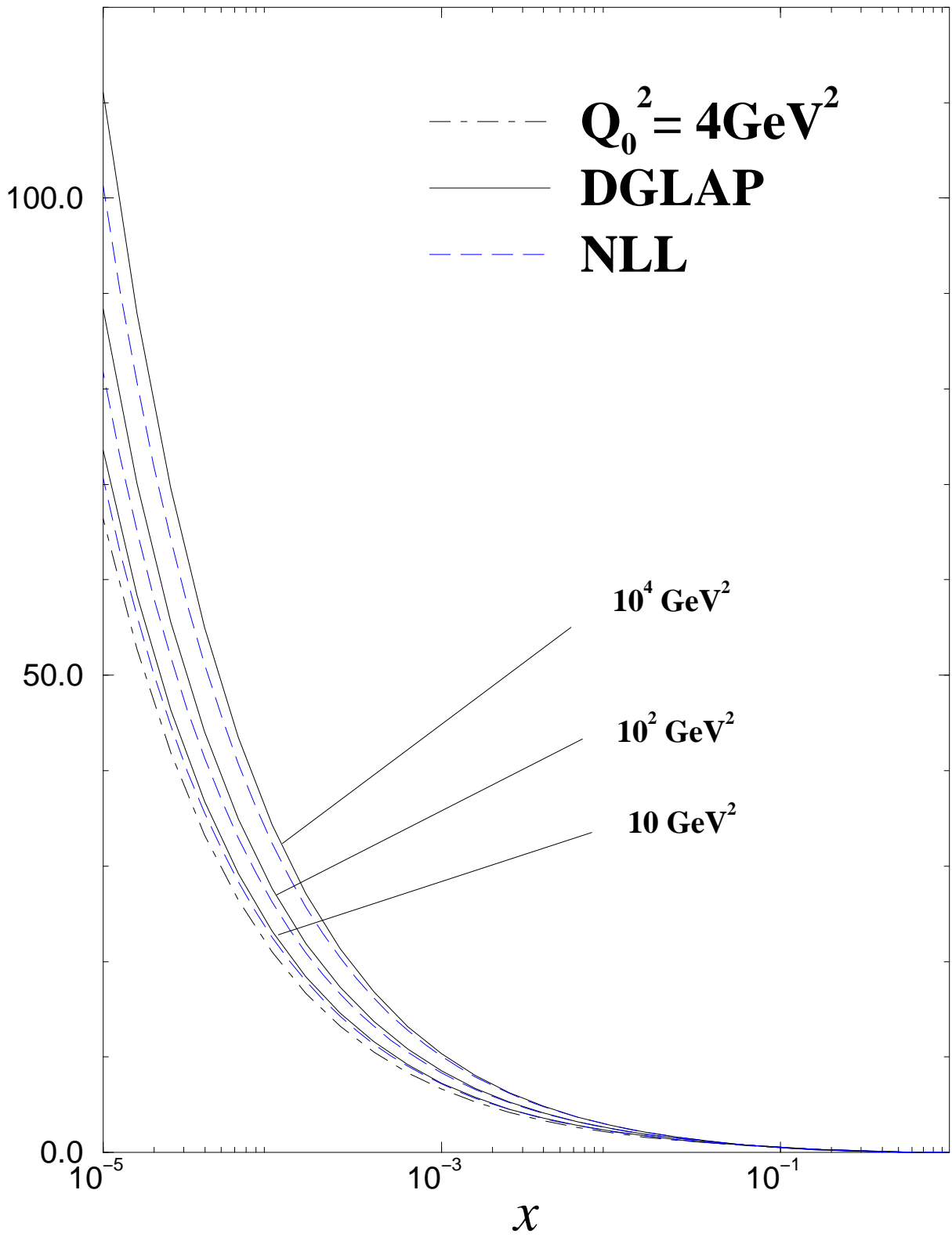
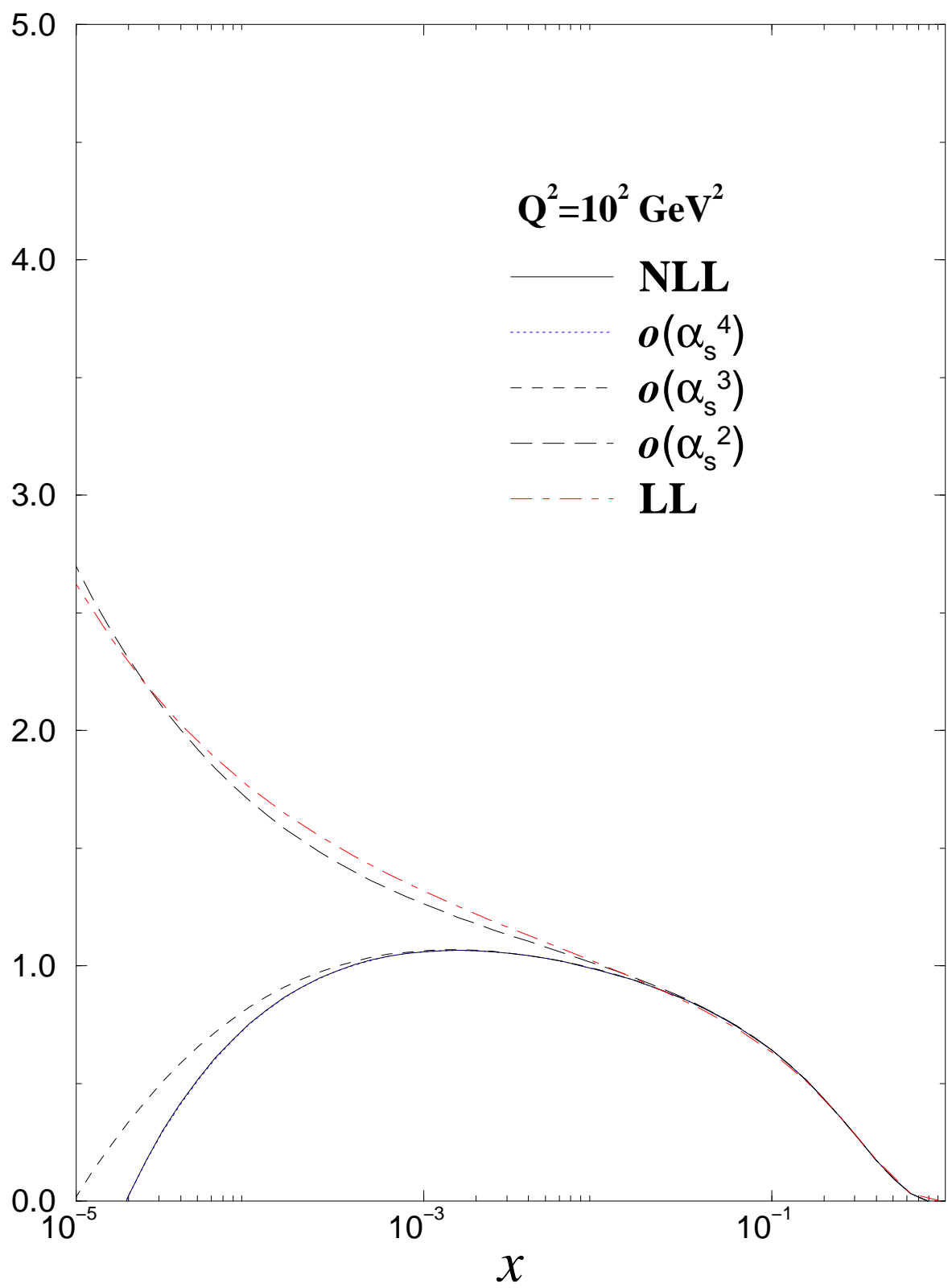
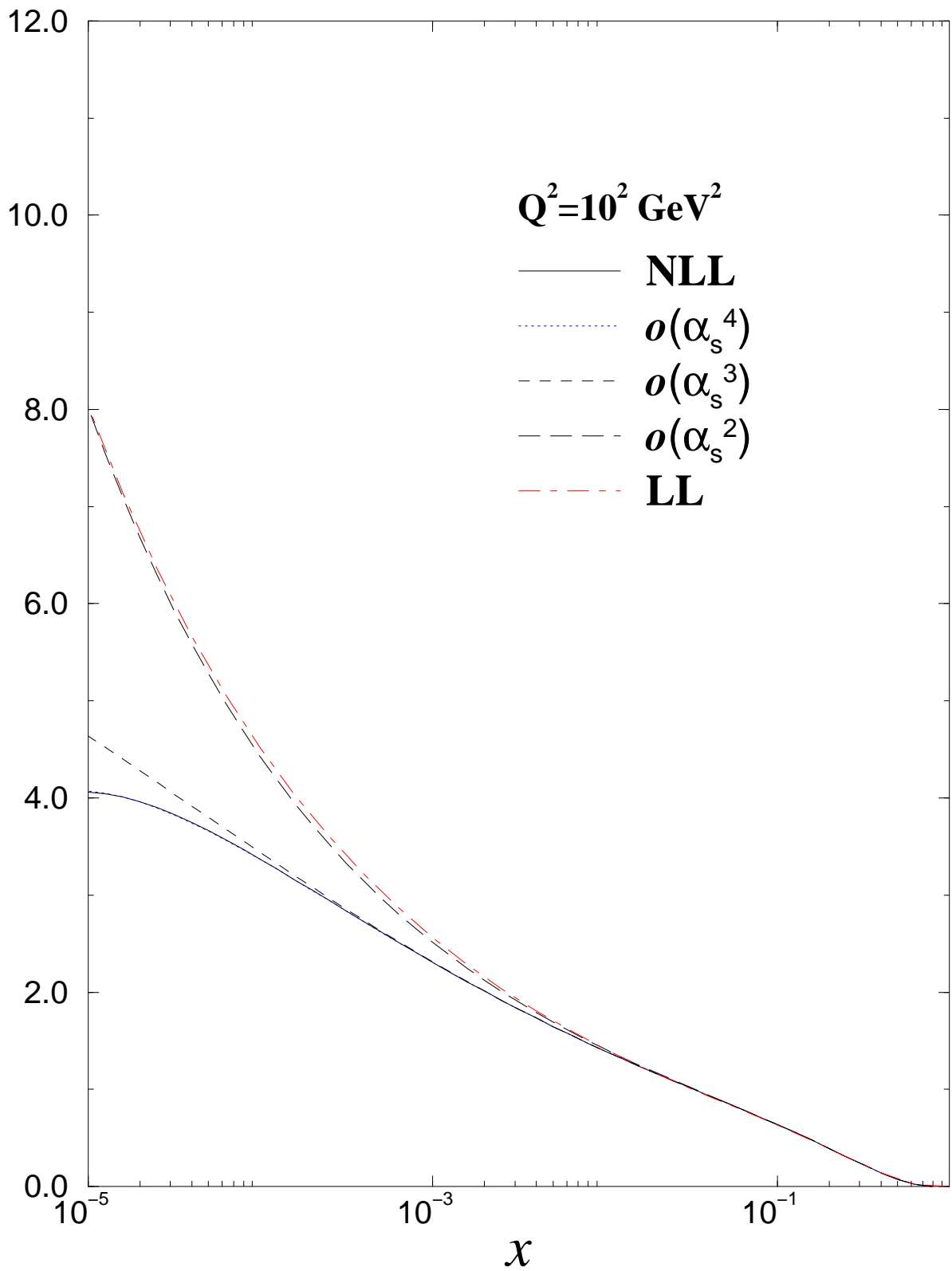
\end{document}